\documentclass[12pt]{article} % Set the font size to 12pt
\usepackage{nips13submit_e,times}
\usepackage{setspace} % Package for controlling line spacing
\usepackage{hyperref}
\usepackage{url}
\usepackage{float}
\usepackage{makecell}
\usepackage{graphicx}
\usepackage{colortbl}
\usepackage[backend=bibtex, style=numeric,sorting=none]{biblatex}
\usepackage{amsmath}
\usepackage{amsfonts}

\usepackage{geometry}
\usepackage{siunitx}
\usepackage{graphicx}
\usepackage{tikz} 
\usepackage{titlesec}
\usepackage{lipsum}
\usepackage{qtree}
\usepackage{multirow}
\usepackage{float}
\usepackage{amsthm}
\usepackage{adjustbox}
\usepackage{amssymb}
\usepackage{amstext}
\usepackage{enumerate}
\usepackage{paralist}
\usepackage{caption}
\usepackage{hyperref}
\usepackage{url}
\usepackage{subcaption}
\usepackage{makecell}
\usepackage{soul}
\usepackage{tabu}
\usepackage{colortbl}
\usepackage[ruled, vlined]{algorithm2e}
\usepackage{algpseudocode}
\usepackage{verbatim}

\title{Personalized graph feature-based multi-omics data integration for cancer subtype identification}

\author{
Saiful Islam\\
Institute for Artificial Intelligence and Data Science, State University of New York at Buffalo\\
 Buffalo, USA\\
Department of Mathematics, University of Dhaka\\
Dhaka, Bangladesh\\
\texttt{saifulis@buffalo.edu} \\
\And
Md. Nahid Hasan\\
Department of Mathematics, Bangladesh University of Engineering and Technology\\
Dhaka, Bangladesh\\
\texttt{nhasan@math.buet.ac.bd}}

\nipsfinalcopy % Uncomment for camera-ready version

\begin{document}

\maketitle

%\begin{abstract}
% I dont think we need an abstract for this submisson
\begin{abstract}
Cancer is a highly heterogeneous disease with significant variability in molecular features and clinical outcomes, making diagnosis and treatment challenging. In recent years, high-throughput omic technologies have facilitated the discovery of mechanisms underlying various cancer subtypes by providing diverse omics data, such as gene expression, DNA methylation, and miRNA expression. However, the complexity and heterogeneity of multi-omics data present significant challenges for their integration in exploring cancer subtypes. Various methods have been proposed to address these challenges. In this paper, we propose a novel and straightforward approach for identifying cancer subtypes by integrating patient-specific subnetworks features from different omics data. We construct patient-specific induced subnetwork using a random walk with restart algorithm from patient similarity networks (PSNs) and compute nine structural properties that capture essential network topology. These features are integrated across the three omic datasets to form comprehensive patient profiles. K-means clustering is then applied for cancer subtype identification. We evaluate our approach on five cancer datasets, including breast invasive carcinoma, colon adenocarcinoma, glioblastoma multiforme, kidney renal clear cell carcinoma, and lung squamous cell carcinoma, for three different omic data types. The evaluation shows that our method produces promising and effective results, demonstrating competitive or superior performance compared to existing methods and underscoring its potential for advancing personalized cancer diagnosis and treatment.
\end{abstract}

\section{Introduction}
%\textcolor{blue}{A description of cancer subtyping and why it is important?}
Cancer is a highly heterogeneous disease, characterized by significant variability in molecular features and clinical outcomes even within the same cancer type. This heterogeneous nature of cancer makes the effective diagnosis and therapy particularly challenging~\cite{zhang2018characterization, alizadeh2015toward, kuijjer2018cancer, ferlay2015cancer}. Researchers have discovered that cancers are comprised of multiple subtypes, each with distinct molecular profiles and clinical outcomes~\cite{curtis2012genomic, kuijjer2018cancer}. In precision medicine, accurately identifying these cancer subtypes is essential for effective patient classification, developing individualized therapeutic strategies, targeted treatment with a high level of efficacy, and improving survival rates~\cite{curtis2012genomic, alizadeh2015toward, kuijjer2018cancer, chin2008translating}. High-throughput omic technologies are increasingly making available various omics data, for example gene expression, DNA methylation and miRNA expression, and these enriched multi-omics data facilitate the discovery of mechanisms underlying various cancer subtypes~\cite{ferlay2015cancer, beroukhim2010landscape}. Combining the analysis of multi-omics data has demonstrated significant power in robust cancer subtype identification compared to analyzing individual omic-data types in isolation~\cite{ritchie2015methods, wang2016integrating}. This comprehensive analysis allows for a broad range of holistic understanding of cancer subtypes and their mechanisms, facilitating the development of effective treatment strategies.

%\textcolor{blue}{Literature review. Methods for cancer subtyping. And why our method is necessary?}
However, the identification of cancer subtypes through the integration and analysis of multi-omics datasets presents significant challenges due to the complexity of these data, characterized by high dimensionality, and heterogeneity~\cite{wang2022network, duan2021evaluation}. Effectively navigating and extracting meaningful insights from such intricate data demands the development of innovative approaches. A plethora of methods have been proposed to integrate multi-omics data for delineating distinct cancer subtypes, tailoring a comprehensive understanding of underlying mechanisms for different subtypes and developing patient-specific therapy~\cite{wang2022network, duan2021evaluation}. Among these methods, probabilistic and network-based approaches are two dominating categories for integrating multi-omics data~\cite{bersanelli2016methods, rappoport2018multi, duan2021evaluation}. Probabilistic methods, for example iCluster~\cite{shen2009integrative} and its variants (iClusterPlus, iClusterBayes), Multi-omics Factor Analysis (MOFA)~\cite{argelaguet2018multi}, LRAcluster~\cite{wu2015fast}, tensorial Independent Component Analysis (tICA)~\cite{beckmann2005tensorial}, and Non-negative Matrix Factorization (NMF)\cite{zhang2012discovery}, learn a joint latent variable to find a single common clustering structure for all involved omics data by assuming that a common latent feature space is shared by different types of omic data. After a low dimensional embedding, the clustering tasks are accomplished through heuristic approaches with a predefined number of clusters. Network-based methods construct similarity networks from different omics data and then integrate these networks for downstream clustering tasks. One of the popular network-based techniques for intergrating multi-omics data is Similarity Network Fusion (SNF)~\cite{wang2014similarity}. SNF first creates a patient similarity network (PSN) for each type of omics data and then fuses all PSNs into a single final network using message passing approach across the PSNs. Another network fusion method based on message passing approach, Affinity Network Fusion (ANF), can incorporate omic-specific weights~\cite{ma2018affinity}. Random walks on multiplex networks have also been used for integrating and analyzing heterogeneous omics data~\cite{wang2016integrating, xu2021network}. An ensemble approach, Perturbation Clustering for Data Integration and Disease Subtyping (PINS), integrates omics data by performing clustering on individual datasets and then combining the clustering results to form connectivity matrices, which are used to identify cancer subtypes~\cite{nguyen2017novel}. Another network-based integration technique, Neighborhood-based Multi-omics Clustering (NEMO), generates the final integrated network by averaging the similarity networks constructed from different omics data~\cite{rappoport2019nemo}. Given the heterogeneity and complexity of omic data, edge weights of similarity networks constructed from different omics data may have different scales and follow different distributions. It is imperative to design a simple yet sophisticated method capable of overcoming the nuances of heterogeneous data landscapes and integrating the networks constructed from these omics data, enabling effective clustering of patients into different cancer subtypes.

In this paper, we propose a novel and straightforward approach for identifying cancer subtypes by integrating patient-specific subnetworks features from different omics data. We leverage three different types of omic data:  mRNA expression, DNA methylation,  and miRNA expression. Our method begins with the construction of  patient similarity network (PSN) for each omic data type using cosine similarity measure. This will emphasize the relationships between similar patients based on their molecular profiles for each omic type. From each omic type PSN, we generate an induced subnetwork for each patient using a random walk with restart algorithm, which explores both neighboring nodes and remote nodes within the PSN. From each of these subnetworks, we compute nine structural properties that capture essential aspects of the network topology. These features are then integrated across the three omic datasets to form a comprehensive patient profile. To identify cancer subtypes, we apply K-means clustering to the aggregated features. We evaluate our method on five benchmark cancer datasets and compare it with four existing methods which are widely used for cancer subtype identification, demonstrating the robustness and effectiveness of our approach. Our approach shows competitive and often superior performance in exploring cancer subtypes, underscoring the potential for advancing personalized cancer diagnosis and treatment.

\section{Materials and Methods}
\begin{figure*}[htbp]
\centerline{
\includegraphics[width=1.01\linewidth]{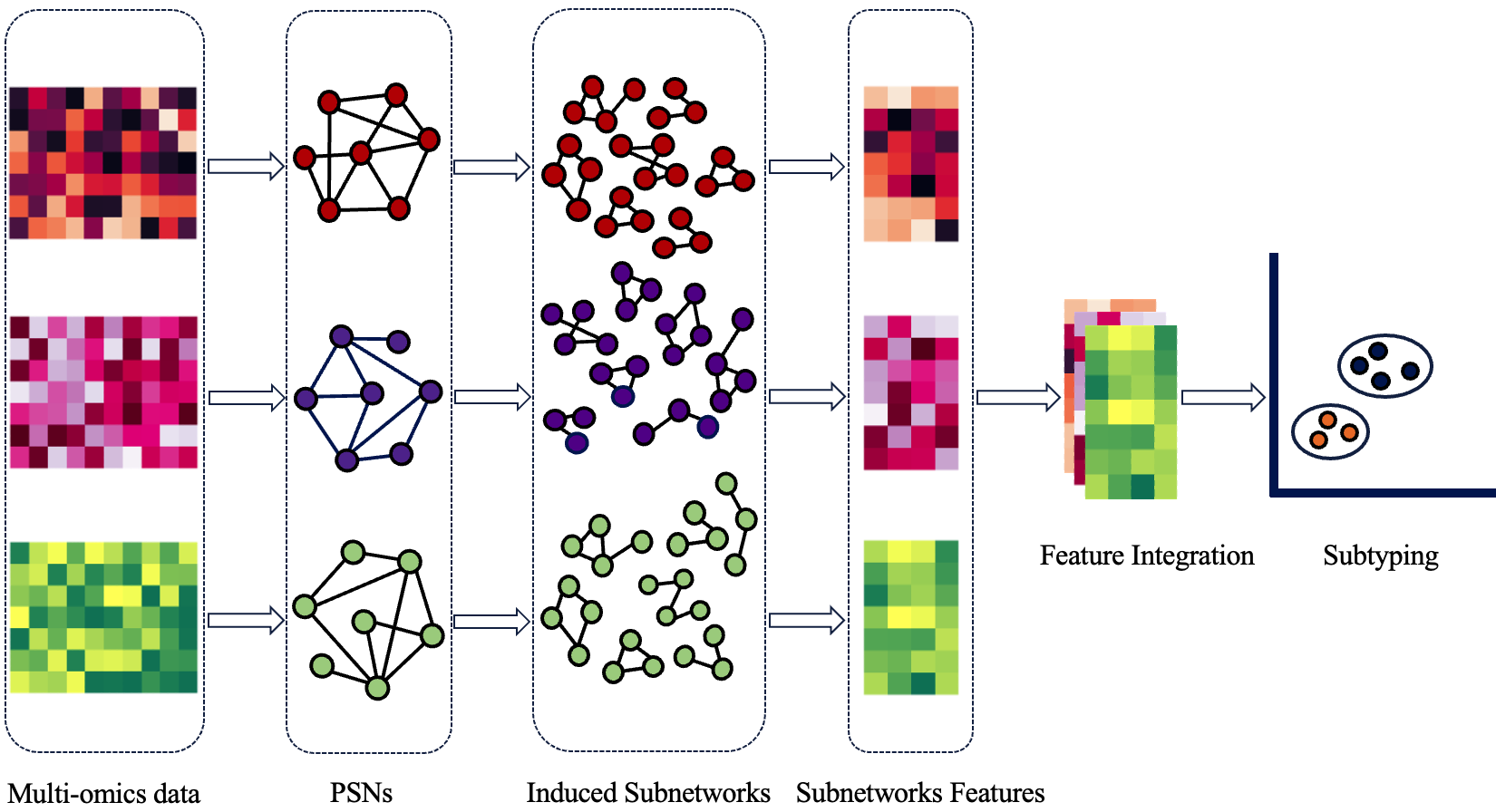}}
\caption{Overview of the proposed approach for identifying cancer subtypes from multi-omics data. First, a Patient Similarity Network (PSN) is constructed from each given omic data. Next, random walk with restart is applied to construct sub-network for each node. Network features are then extracted from these sub-networks. Subsequently, these network features are aggregated. Finally, K-means algorithm is employed to identify cancer subtypes.}
\label{fig:scheme}
\end{figure*}

The methodology of our proposed approach is outlined schematically in Fig.~\ref{fig:scheme}. This figure provides an overview of the comprehensive workflow designed to integrate and analyze multi-omics data for cancer subtype identification. In the following subsections, we elaborate on each step of the approach.

\subsection{Patient Similarity Network}
Our proposed approach directly uses the network features of each subject from each type of omics data. For a given omic data $X^{(m)}$, where $m$ denotes the omic type, we construct a patient similarity network (PSN) with adjacency matrix $A^{(m)}$. This adjacency matrix is computed based on the cosine similarity between the samples. The edge weight between samples $i$ and $j$ is computed using a similar procedure has been used in NEMO~\cite{rappoport2019nemo}, given by the following equation.
\begin{equation}
A^{(m)}_{ij}= \frac{\text{sim}(i,j) I(j\in N_i)}{\sum_{k\in N_i}  \text{sim}(i,k)} +\frac{\text{sim}(i,j)I(i\in N_j)}{\sum_{k\in N_j}  \text{sim}(j,k)}
\label{net_edge}
\end{equation}
where $I(j\in N_i)$ is an indicator function that is 1 if $j$th sample is in the $k-$nearest neighbor of the sample $i$, and otherwise 0. and  $\text{sim}(i,j) $ is defined as bellow.
\begin{equation}
\text{sim}(i,j) =1+ \frac{\langle x^{(m)}_i, x^{(m)}_j \rangle}{ ||x^{(m)}_i||\, ||x^{(m)}_j||}.
\label{sim}
\end{equation}
Equation~\eqref{sim} represents the transformed cosine similarity between the samples $i$ and $j$ which ranges from 0 to 2. This transformation ensures that the edge weights are non-negative. In the Eq.~\eqref{sim}, $\langle, \rangle$ denotes the inner product, $||.||$ represents the $L2$ norm, and $x^{(m)}_i$ is the feature vector of the $i$th sample in the $m$th-type of omics data. 

\subsection{Subnetwork Construction}
From the PSN for each type of omics data, we construct a subnetwork for each sample. To construct a subnetwork for a given sample, we follow a random walk with restart procedure. We initiate the random walk from a node, and the walker chooses an edge according to the weights of the edges, subsequently moving to its neighboring node. Additionally, at each step, the walker has a positive probability of returning to the starting node. In this study, we perform ten random walks per node, each with a length of six steps, and set the restart probability to 0.15. The random walks collect nodes, and we then construct a subnetwork induced by the set of unique nodes collected during these walks.

\subsection{Subnetwork Features}
For each sample, we computed nine global network features from the subnetwork. These features encompass various aspects of network topology and connectivity, including the mean degree of the nodes, mean node strengths, coefficient of variation (CV) of node strengths, weighted density, the trace, the largest eigenvalue, and the second largest eigenvalue of the Laplacian matrix, mean clustering coefficients, mean weighted betweenness centrality, and mean weighted closeness centrality. Our underlying assumption is that samples belonging to the same cancer subtype exhibit similar network structures, thus resulting in similar global features. Consequently, these extracted features serve as discriminative quantities in the network feature space, aiding in the identification of distinct cancer subtypes. In a previous study on graph classification tasks, we used similar types of network features and achieved high-level classification accuracy~\cite{islam2024structural}. The rationale for using these features is based on their effectiveness in capturing critical structural characteristics of networks. 
By leveraging these features, we aim to ensure that our analysis is grounded in proven approaches that have demonstrated their power in  graph classification tasks.

\subsection{Network Features Fusion}
For the $n$ samples and $m$-type omics data, we have $n\times d$, $m$ network feature matrices $\tilde{X}^{(1)}, \tilde{X}^{(2)}, \cdots, \tilde{X}^{(m)}$, where $d=9$. For the multi-omics feature fusion, we aggregate these matrices by taking their average. The aggregated feature matrix is given by
\begin{equation}
X_{agg} = \frac{1}{m} (\tilde{X}^{(1)}+\tilde{X}^{(2)}+ \cdots +\tilde{X}^{(m)}).
\label{agg_feature}
\end{equation}
Therefore, the dimension of the aggregated feature vector for each of the samples is nine, which is much lower than the original feature dimension. 

\subsection{Clustering the Samples}
After aggregation, we obtain a low dimensional representation for each sample. For identifying cancer subtypes, we employ K-means algorithm to obtain clusters. We determine the optimal number of clusters by using silhouette score~\cite{rousseeuw1987silhouettes}. This score provides the quality of clusters by computing the compactness within clusters and the separation between different clusters. The silhouette score for a sample $i$ is given by 
\begin{equation}
s_{i} = \frac{b(i)-a(i)}{\text{max} \{a(i), b(i)\}},
\label{silhoutte}
\end{equation}
where $a(i)$ is the average distance of all the points within the same cluster to which the sample $i$ belongs to and $b(i)$ represents the smallest average distance of the sample $i$ to all points in any other clusters. We then average the silhouette score across all the samples. This score ranges from $-1$ to $1$, with higher values indicating that the sample is well matched to its own cluster.

\subsection{Data Acquisition and Analysis}
%prostate adenocarcinoma (PRAD), (2) pancreatic adenocarcinoma (PAAD),(3) lung adenocarcinoma (LUAD), (4) lung squamous cell carcinoma (LUSC), (5) kidney renal papillary cell carcinoma (KIRP), (6) kidney renal clear cell carcinoma (KIRC), (7) glioblastoma multiforme (GBM), (8) colon adenocarcinoma (COAD), and (9) breast invasive carcinoma (BRCA).
In this study, we assess the performance of our proposed approach using five datasets sourced from The Cancer Genome Atlas (TCGA). Each dataset comprises multi-omics data encompassing mRNA expression, DNA methylation,  and miRNA expression formats. The datasets are used in our analysis include breast invasive carcinoma (BIC),  colon adenocarcinoma (COAD), glioblastoma multiforme (GBM), kidney renal clear cell carcinoma (KRCCC), and lung squamous cell carcinoma (LSCC). Prior to analysis, we perform standard preprocessing steps. First of all, samples with more than 20\% missing features are removed, followed by the elimination of features with missing values exceeding 20\% of the samples. In order to impute the remaining missing values, the k-nearest neighbors (k-NN) method is employed. Subsequently, each feature is normalized across all samples to ensure uniformity in scale. Following preprocessing, the dataset characteristics are as follows: BIC contains 105 samples with 17,814 mRNA expression features, 23,094 DNA methylation features, and 354 miRNA expression features; COAD comprises 92 samples with 17,814 mRNA expression features, 23,088 DNA methylation features, and 312 miRNA expression features; GBM comprises 215 samples with 12,042 mRNA expression features, 1,305 DNA methylation features, and 534 miRNA expression features;  KRCCC includes 122 samples with 17,899 mRNA expression features, 24,960 DNA methylation features, and 329 miRNA expression features; and LSCC consists of 106 samples with 12,042 mRNA expression features, 23,074 DNA methylation features, and 352 miRNA expression features.

\section{Results and Discussion}
\begin{figure}[htbp]
\centerline{
\includegraphics[width=.65\linewidth]{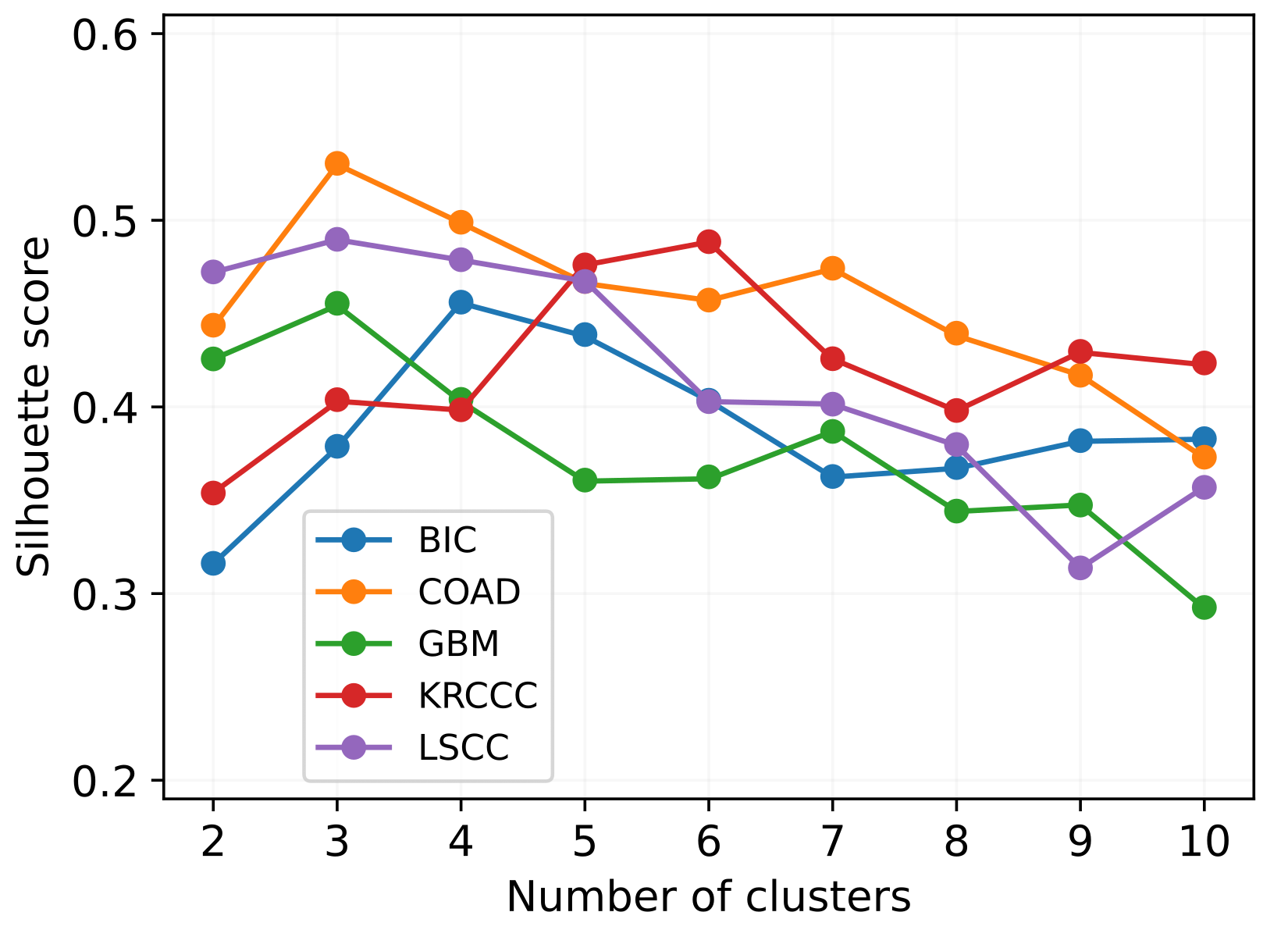}}
\caption{Average silhouette scores for the number of clusters, $K\in \{2,3,\cdots, 10\}$ across five datasets.}
\label{fig:silhouette}
\end{figure}

We employed the K-means clustering algorithm to identify cancer subtypes for $K\in \{2,3,\cdots, 10\}$. We determined the optimal number of subtypes by utilizing the average silhouette score. Fig.~\ref{fig:silhouette} shows the average silhouette scores for $K\in \{2,3,\cdots, 10\}$. Our results  indicate that the optimal number of subtypes for BIC, COAD,  GBM,  KRCCC, and LSCC are four, three, three, six, and three, respectively. 

\begin{figure}[htbp]
\centerline{
\includegraphics[width=0.56\linewidth]{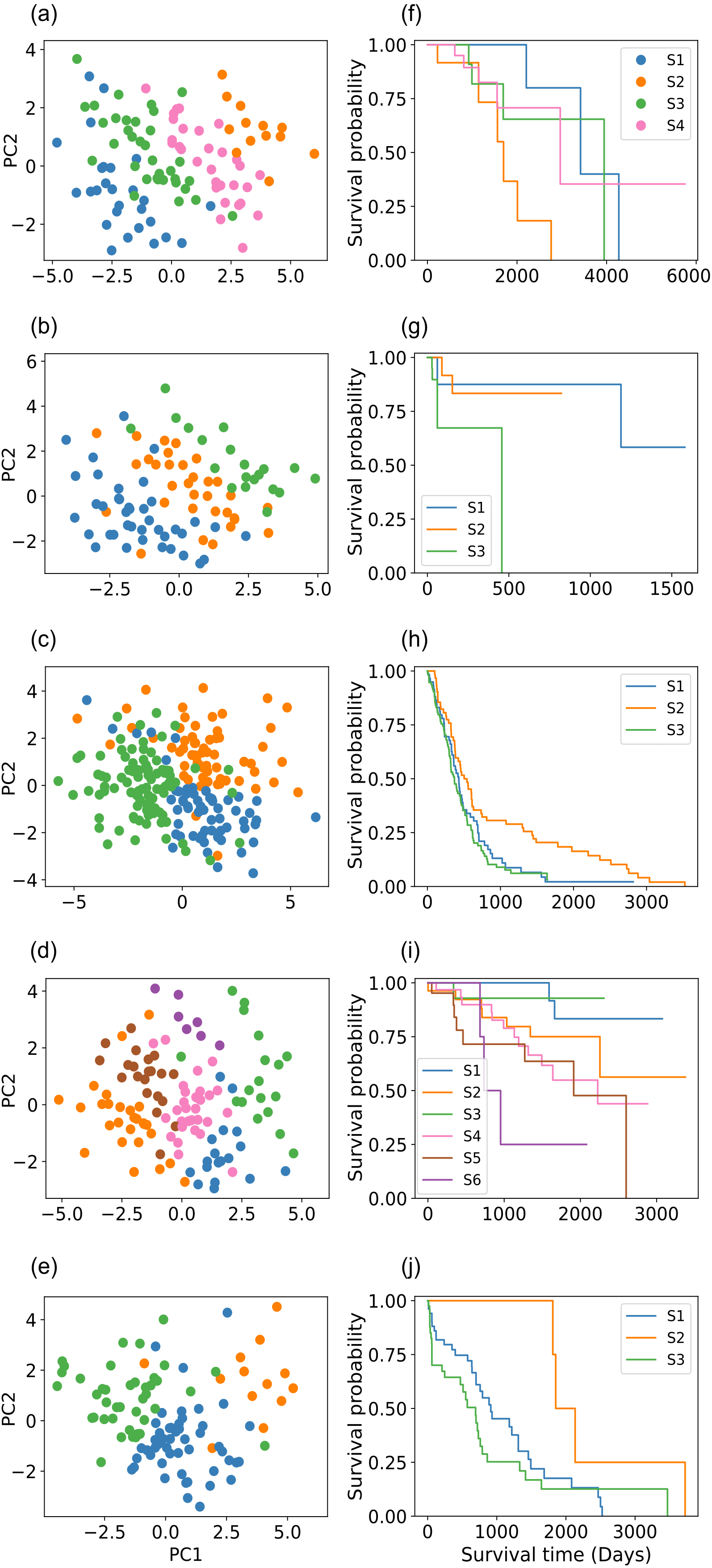}}
\caption{Survival analysis curves of the patient subtypes and two-dimensional representations of aggregated features for five datasets. Panels (a)-(e) display the two-dimensional representations of the aggregated features for individual patients in the datasets: (a) BIC, (b) COAD, (c) GBM, (d) KRCCC, and (e) LSCC. Each circle represents a patient. Panels (f)-(j) demonstrate the survival curves corresponding to the datasets BIC, COAD, GBM, KRCCC, and LSCC, respectively. The patient subtypes are denoted as S1, S2, S3, S4, S5, and S6, with the colors in the survival curves indicating the respective subtypes. The $-\log10$ of the $p-$values in the Cox log-rank test are 0.0101, 0.0083, 0.0039, 0.0016, and 0.0039 for BIC, COAD, GBM, KRCCC, and LSCC, respectively.}
\label{fig:survival}
\end{figure}

To visualize different subgroups of patients, we applied Principle Component Analysis (PCA) to the aggregated nine-dimensional feature vectors. The first two principal components (PC1 and PC2) for each dataset are presented in the first column of Fig.~\ref{fig:survival}. The Fig.~\ref{fig:survival}~(a)--(e) are the PC1 and PC2 visulaization for BIC, COAD,  GBM,  KRCCC, and LSCC, respectively. Different subtypes are indicated with different colors. We observe clear distinctions between subtypes even in this two-dimensional representation. This distinct grouping in a reduced two-dimensional space provides a reasonable evidence that the underlying nine-dimensional feature set captures significant variation and relevant information necessary for distinguishing between different cancer subtypes.

We conducted survival analysis using the Kaplan–Meier method~\cite{bland1998survival}, which estimates survival probabilities over time. To evaluate the statistical significance of the differences between cancer subtypes, we used the log-rank test within the Cox regression framework~\cite{hosmer2008applied}. The results of the survival analysis are shown in the second column of Fig.~\ref{fig:survival}. The Fig.~\ref{fig:survival}~(f)--(j) show the survival curves for different subtypes in BIC, COAD,  GBM,  KRCCC, and LSCC, respectively. The p-values, we obtained are 0.0101, 0.0083, 0.0039, 0.0016, and 0.0039 for BIC, COAD,  GBM,  KRCCC, and LSCC, respectively. These results indicate statistically significant differences between the cancer subtypes across five datasets with a significance level of 0.05. 
The statistical significance of the differences between cancer subtypes underscores the effectiveness of our approach in distinguishing between cancer subtypes. This also suggests that the aggregated nine-dimensional features obtained from subnetworks of each patients from three different multi-omics data are robust and capable of capturing meaningful differences. 
The significant p-values across various cancer types further highlight the potential utility of our proposed approach in precision medicine for accurate classification and stratification of cancer patients.

%\begin{figure*}[htbp]
%\centerline{
%\includegraphics[width=1.01\linewidth]{All_PC_survival_plot.png}}
%\caption{Survival analysis curves of patient subtypes and two-dimensional representations of aggregated features for five datasets. Panels (a)-(e) display survival curves for the datasets: (a) BIC, (b) COAD, (c) GBM, (d) KRCCC, and (e) LSCC. The subtypes are denoted as S1, S2, S3, S4, S5, and S6. Panels (f)-(j) present the two-dimensional representations of the aggregated features of patients for the datasets BIC, COAD, GBM, KRCCC, and LSCC, respectively, using PC1 and PC2. Each circle represents a patient, with colors corresponding to the subtypes indicated in the survival curves.}
%\label{fig:survival}
%\end{figure*}

We compared the performance of our approach with four state-of-the-art methods for cancer subtyping using multi-omics datasets. The methods include iClusterPlus~\cite{mo2013pattern}, LRAcluster~\cite{wu2015fast}, SNF~\cite{wang2016integrating}, and PINS\cite{nguyen2017novel}. For the SNF, we used  the published results from their paper, as the same preprocessed datasets with an identical number of samples are used in our study. For the remaining methods, we generated results using the parameter settings specified in the respective papers. A comparative analysis of the p-values from the survival analysis is presented in Table~\ref{tab1}. The results show that our method outperforms other methods in the KRCCC dataset. Moreover, our method demonstrates superior performance compared to iClusterPlus, LRAcluster, and PINS in the COAD and GBM datasets; it surpasses iClusterPlus, LRAcluster, and SNF in the LSCC dataset; and it performs better than LRAcluster and PINS in the BIC dataset. These findings highlight the effectiveness and robustness of our approach in identifying cancer subtypes. This comparison underscores the potential advantages of our approach in integrating multi-omics data for accurate cancer subtyping.

\begin{table}[htbp]
\caption{Performance comparison on five datasets. In each cell S/P represents optimal number of subtypes/p-values in log-rank test, with bold numbers indicating significant results.  %\textcolor{blue}{We need to complete this table. I will send you the code for NEMO, please run it and let me know if it works on your computer.}
}
\begin{center}
\begin{tabular}{cccccc}
\hline
Dataset& BIC & COAD & GBM & KRCCC &LSCC\\
Method & & & & & \\
\hline
iClusterPlus & 4/\textbf{0.0008} &2/0.3272 & 4/\textbf{0.0240 }& 5/\textbf{0.0024}& 4/\textbf{0.0088} \\
LRAcluster & 4/\textbf{0.0385} & 4/\textbf{0.01681} & 5/0.31103& 5/0.05272 & 3/\textbf{0.0078} \\
SNF & 5/\textbf{0.0011}& 3/\textbf{0.0008} &3/\textbf{0.0002} &3/\textbf{0.0290} & 4/\textbf{0.0200} \\
%CC & & & & & \\
%NEMO & & & & & \\
PINS& 3/\textbf{0.0307} &8/\textbf{0.0104 }&2/\textbf{0.01044} &2/0.0562 & 4/\textbf{0.0030}\\
%ANF & & & & & \\
Our  results & 4/\textbf{0.0101}& 3/\textbf{0.0083 }& 3/\textbf{0.0039} & 6/\textbf{0.0016} & 3/\textbf{0.0039}\\
%\hline
%
%\hline
%SNF & 5/0.0011& 3/0.00088&3/0.0002 &3/0.029 & 4/0.02 \\
%CC & & & & & \\
%%NEMO & & & & & \\
%iCluster+ & 5/0.00626 &3/0.19073 & 4/0.0240 & 5/0.00085 & 4/0.00886 \\ %krccc 3/0.00368, 4/0.00164
%PINS& 3/0.03077 &8/0.01044 &2/0.01044 &2/0.05629 & 4/0.00296\\
%%ANF & & & & & \\
%Our  & 4/0.0101& 3/0.0083 & 3/0.0039& 6/0.0016 & 3/0.0039\\
\hline
\end{tabular}
\label{tab1}
\end{center}
\end{table}

We also compare the subtypes identified by our approach with four established subtypes: Basal-like, HER2, Luminal A, and Luminal B for the BIC~\cite{cancer2012comprehensive}. The comparison results are provided in Table~\ref{pam}. We found that the   Subtype 1 is enriched for Luminal A, and Basal-like and HER2 are over presented in the Subtype 3.
\begin{table}[htbp]
\caption{Comparison of the identified subtypes with known subtypes for BIC
%on five datasets. Total samples is 105, but in the table we have 104, because in PAM50 one sample has no type.
}
\begin{center}
\begin{tabular}{cccccc}
\hline
& Basal-like & HER2 & Luminal A & Luminal B \\%& Normal\\
\hline
  Subtype 1   &  4  &  0  & 20 &  2  \\%& 1\\
  Subtype 2   &  2  &  3  & 2   & 5  \\%& 0\\
  Subtype 3   & 12 &  17 & 3 &  3  \\%& 0\\
  Subtype 4   &  5  &  5  & 17 &  2  \\%& 1\\
 \hline
\end{tabular}
\label{pam}
\end{center}
\end{table}

%\begin{table}[htbp]
%\caption{Performance comparison on five datasets.}
%\begin{center}
%\begin{tabular}{ccccc}
%\hline
%Subtype  & Classical-like &Mesenchymal  &Neural  & Proneural\\
% \hline                              
%  Subtype 1	&14           &18       &7         &10\\
%  Subtype 2	&13           &13       &3         &25\\
%  Subtype 3	& 21          &22      &17         &17\\
% \hline
%\end{tabular}
%\label{pam}
%\end{center}
%\end{table}

Moreover, the comparison of our results with four established subtypes of GBM: Classical-like, Mesenchymal, Neural, and Proneural~\cite{ceccarelli2016molecular} presented in Table~\ref{tcga-gbm}. This comparison showed that the Subtype 1 is enriched for Mesenchymal, and Subtype 3 spread across all the known subtypes but enriched with Classical-like, and Mesenchymal.

\begin{table}[htbp]
\caption{Comparison of the identified subtypes with known subtypes for GBM}
\begin{center}
\begin{tabular}{ccccc}
\hline
Subtype  & Classical-like &Mesenchymal  &Neural  & Proneural\\
 \hline                              
  Subtype 1	&17           &22       &7         &12\\
  Subtype 2	&14           &13       &11         &12\\
  Subtype 3	& 27          &31      &16        &15\\
 \hline
\end{tabular}
\label{tcga-gbm}
\end{center}
\end{table}

\section{Conclusion}
In this paper, we proposed a novel and straightforward approach for identifying cancer subtypes by integrating patient-specific subnetwork features from multi-omics data. For three different types of omics data, mRNA expression, DNA methylation, and miRNA expression, we constructed PSNs and then employed a random walk with restart algorithm to generate induced subnetworks for each patient within the PSNs. By computing nine structural properties for each subnetwork, which capture essential aspects of network topology, we formed feature vectors for each patient that provide comprehensive profiles in the network properties space. The K-means clustering has been applied to the integrated features to obtain cancer subtypes and the optimal number of clusters has been determined by using the silhouette score. We evaluated the performance of our approach on five cancer datasets and compared it with four existing methods. The results demonstrated that our method produced promising and effective outcomes, often outperforming existing methods in terms of p-values in Cox regression analysis. Additionally, we showed that the identified subtypes had enrichment relations with the known subtypes in the cases of breast invasive carcinoma and glioblastoma multiforme. These findings underscore the robustness and potential of our approach for advancing personalized cancer diagnosis and treatment. Our method addresses the challenges posed by the complexity and heterogeneity of multi-omics data, providing a straightforward yet sophisticated solution for cancer subtype identification. Future work will extend the study to other cancer types and incorporate additional omics data with rigorous biological and clinical analysis. Furthermore, we aim to identify the most effective network features of subnetworks across omics data to further improve our method's performance in subtype identification and enhance its scalability.

\printbibliography

\end{document}